# Predicting Superconducting Transition Temperature through Advanced Machine Learning and Innovative Feature Engineering

Hassan Gashmard, Hamideh Shakeripour, Mojtaba Alaei

*Department of Physics, Isfahan University of Technology, Isfahan 84156-83111, Iran*

Corresponding author: Hamideh Shakeripour (hshakeri@iut.ac.ir)

**Abstract**

Superconductivity is a remarkable phenomenon in condensed matter physics, which comprises a fascinating array of properties expected to revolutionize energy-related technologies and pertinent fundamental research. However, the field faces the challenge of achieving superconductivity at room temperature. In recent years, Artificial Intelligence (AI) approaches have emerged as a promising tool for predicting such properties as transition temperature ($T_c$) to enable the rapid screening of large databases to discover new superconducting materials. This study employs the SuperCon dataset as the largest superconducting materials dataset. Then, we perform various data pre-processing steps to derive the clean DataG dataset, containing 13022 compounds. In another stage of the study, we apply the novel CatBoost algorithm to predict the transition temperatures of novel superconducting materials. In addition, we developed a package called Jabir, which generates 322 atomic descriptors. We also designed an innovative hybrid method called Soraya package to select the most critical features from the feature space. These yield $R^2$ and RMSE values (0.952 and 6.45 K, respectively) superior to those previously reported in the literature. Finally, as a novel contribution to the field, a web application was designed for predicting and determining the $T_c$ values of superconducting materials.

## 1. Introduction

The amazing properties of superconducting materials are a direct consequence of quantum mechanics that emerge on a large scale [1]. The two basic characteristics of superconductors that make this class of materials different from others include: a) offering no resistance to the flow of electric currents, and b) complete exclusion of magnetic field [2]. No comprehensive theory capable of predicting the transition temperatures ($T_c$) of superconducting materials has yet been presented to date and the discovery of new superconductors still relies on expert intuition and is largely dependent on trial and error based on experience [3]. Hence, empirical laws have for many years served as guides for researchers in their efforts to fabricate new superconducting materials [4].

Condensed Matter Physics strives to discover the interactions of materials at the atomic level since material properties are derived from these interactions [5]. Prediction and determination of the microscopic properties of materials presuppose the solution of the Schrodinger equation for a Many-Body system. However, solving this equation for such systems is practically impossible due to the vast Hilbert space needed to handle them, especially for highly correlated materials. Consequently, a solution adopted in most cases is to employ approximate methods [6-8]. One of these methods is Density Functional Theory (DFT) which is based on the Hohenberg-Kohn and Kohn-Sham theorems and has a substantial record of success in predicting material properties and solving the associated quantum mechanics problems [9-11]. Despite its outstanding achievements, the theory has some limitations in its current form; for instance, it employs approximation for exchange-correlation functional, yields errors when used for strong correlation systems, can only be employed for a small number of atoms, and is hampered by increasing computational costs and runtime with increasing system size [10, 12-15]. Strong electron-electron correlations in superconducting materials make it extremely challenging to perform first-principles calculations to determine their structural properties and predict their $T_c$ [3, 9], making searching for novel alternative approaches inevitable.

As alternative strategies for solving quantum mechanics problems, machine learning methods offer lower computation costs, shorter execution times, accurate predictions, and faster development cycles [9, 12, 13]. Being data-driven and given the fact that huge amounts of data have been produced over the years, machine learning methods encourage researchers to utilize them for discovering novel materials and predicting their properties [4-6, 16]. Materials Science is nowadays said to have entered its fourth stage of evolution, termed "Data-Based Materials Science", a term borrowed from Thomas Samuel Kuhn to describe the field's development [6, 12, 17, 18]. Figure 1 illustrates the four (empirical, theoretical, computational, and data-driven) paradigms of materials science. To date, large amounts of theoretical and experimental data have been collected in the three traditional (*i.e.*, empirical, theoretical, and computational) paradigms; the next step, logically, is to apply the new innovative tools developed by artificial intelligence, which are capable of extracting knowledge from such data [6, 12, 18-22].

Given the importance of the $T_c$ values of superconducting materials, researchers have in recent years developed machine learning-based models for predicting this quantity. Selecting 21,263 superconducting materials and utilizing 80 atomic descriptors for each compound, Hamidieh [4] used the XGBoost algorithm to design a model for predicting of $T_c$. Stanev *et al*. [16] employed the Random Forest algorithm to develop a model using 132 atomic features of Magpie descriptors for 6196 superconducting compounds. Konno *et al*. [3] implemented a convolutional neural network (CNN) model (*i.e.*, a deep learning model) to predict the $T_c$ values of about 13,000 superconducting materials. They represented their materials using an innovative "periodic table reading" method. The dimensions of the representation were $4 \times 32 \times 7$, with 4 representing the four orbitals of s, p, d, and f corresponding to the valence electrons of each element in a compound, and 32 and 7 denoting the dimensions of the periodic table. Dan *et al*.[23] developed the ConvGBDT model by merging the convolutional neural network (CNN) and the gradient boosting decision tree (GBDT) models. For the three datasets of DataS, DataH, and DataK, the authors used the Magpie descriptors to represent materials and the ConvGBDT model to predict $T_c$ values. Li *et al*. [11] introduced a hybrid neural network (HNN) model as a combination of a convolutional neural network (CNN) and a long short-term memory neural network (LSTM). They utilized atomic vectors and employed both the one-hot and Magpie material characterization methods to represent superconductors in the feature space. The authors found that the Magpie features generally outperformed the one-hot features. Roter *et al*. [24] employed the Bagged Tree method (a variant of the Random Forest algorithm) to design a model for predicting $T_c$. They represented superconducting materials using a chemical composition matrix as the feature space. The matrix had about 30,000 rows and 96 columns, wherein each row corresponded to a chemical formula, and the columns contained the 96 primary elements of the periodic table. Each entry in this matrix was filled with an index corresponding to the elements of each chemical compound. Quinn *et al*. [25] utilized a Crystal Graph Convolutional Neural Network (CGCNN) model to integrate classification and regression models within a pipeline to identify candidates of high-temperature superconductors from among the 130,000 compounds in the Materials Project. In the crystal-graph representation of materials, the connections between atoms represent the graph's edges, and the locations of the atoms and their properties represent the vertices.

The main objective of the current research is to design a suitable and reliable model for predicting the $T_c$ values of superconducting materials using machine learning approaches. While the algorithm and the dataset are the two indispensable research tools in data science, the present study attaches more importance to the dataset than the algorithm. After carefully cleaning data, we generate a suitable feature space for superconducting materials. The main advantages of the present work over previous ones include: 1) Establishing more appropriate feature space related to superconducting $T_c$ and 2) Identifying the features most related to the $T_c$ values of superconducting materials. We reach significant results by designing the Jabir package to produce 322 atomic features for each compound and Soraya package for selecting features.

## 2. Data

### 2.1. Data Set

Two essential steps must be taken before statistical learning can predict $T_c$ in superconducting materials. The first involves collecting and preprocessing a dataset, and the second is adopting a suitable algorithm for the learning process and model development on that dataset. According to Halevy *et al*. [26], the first step is of greater significance as data scientists typically devote about 80% of their efforts to datasets and their preprocessing [27]; the same is valid with the present work using SuperCon dataset (https://doi.org/10.48505/nims.3739), currently the largest and most comprehensive superconducting materials database containing 33407 superconducting compounds.

Here, a significant contribution is done by executing distinct steps of data pre-processing and providing detailed explanations for each step. Ultimately, following the implementation of various data pre-processing phases and the exclusion of problematic data, DataG dataset consisting of 13,022 superconducting compounds is derived.

### 2.2. Cleaning the dataset

**A.** *Dealing with missing and duplicated data*

The SuperCon dataset lacks the transition temperature values for 7088 compounds. These cases are identified as missing data and removed from the dataset. Along with that, we remove 7418 data duplications. Among these, 1264 compounds are regarded as duplicates due to the displacement of data elements; examples include: $MgB_2$, $B_2Mg$, $Ag_7B_1F_4O_8$, $Ag_7F_4O_8B_1$, $Al_{0.1}Si_{0.9}V_3$, $V_3Si_{0.9}Al_{0.1}$, $Zr_2Co_1$, $Co_1Z_2$, . . .

**B.** *Dealing with problematic data*

1) We eliminate 5348 compounds whose element subscripts are X, Y, Z, D, x, y, z, and d. 2) We remove

problematic compounds such as: $HgSr_2Ho_{0.333}Ce_{0.667}Cu_2O_{6=z}$, $Ba_2Cu_{1.2}Co_{2.4}O_{2,4}$, $Ag_7Bf_4O_8$, $Hg_{0.3}Pb_{0.7}Sr_{1.75}La_{0.25}CuO_{4+2}$, $Ho_{0.8}Ca_{0.2}Sr_2Cu_{2.8}P_{0.2}O_{z+0.8}$, $Bi_{1.6}Pb_{0.4}Sr_2Ca_2Cu_3F_{0.8}O_{z-0.8}$. 3) Compounds containing the elements not included in the periodic table are ignored. 4) Given the objective of predicting transition temperatures for superconductors at ambient pressure, those created under non-ambient pressures (e.g., $La_1H_{10}$, $H_2S_1$, $H_3S_1$, $D_3S_1$, ...) are removed from the dataset. 5) The compound $YBa_2CuO_{6050}$ is eliminated on the grounds that the oxygen subscript of 6050 might be incorrect [4]. 6) We dismissed 70 compounds whose transition temperatures are reported to be zero. 7) Finally, the compound $Pb_2CAg_2O_6$ is discarded due to the unreasonable transition temperature of 323 K reported for this compound.

**C.** *Data correction*

1) According to the SuperCon reference [28], the transition temperature of the iron-based superconductor $CsEuFe_4As_4$ is nearly 30 K, while the SuperCon dataset records it as 287 K. Therefore, it is modified to 28.7 K. Moreover, the compound $Sm_1Ba_{-1}Cu_3O_{6.94}$ is substituted with $Sm_1Ba_1Cu_3O_{6.94}$. 2) $Bi_{1.6}Pb_{0.4}Sr_2Cu_3Ca_2O_{1013}$ is altered to $Bi_{1.6}Pb_{0.4}Sr_2Cu_3Ca_2O_{10.13}$ because the nearby data rows containing formulas with $O_{10.xx}$ [4].

**D.** *Dealing with multiple temperatures reported for a single compound*

One limitation in the SuperCon dataset is the presence of multiple $T_c$ values reported for 2132 compounds, posing a challenge for accurate analysis. For instance, $MgB_2$ alone has been reported to exhibit 47 different transition temperatures ranging from 5 K to 40.5 K. To tackle this challenge, it has been recommended to consider average transition temperatures for compounds that have multiple $T_c$ values reported in the dataset. Prior to determining the average $T_c$ value, it is essential to exclude compounds whose reported transition temperatures display significant dispersion. To achieve this, the standard deviation of the different transition temperatures for each compound is calculated and compounds with standard deviations greater than 20 K are removed from the dataset. Performing this procedure leads to the elimination of 18 compounds.

**E.** *Detecting outliers*

Undoubtedly, outliers in a dataset can pose problems in identifying underlying patterns, resulting in diminishing system performance and accuracy [29]. In this study, the outlier data are detected using the Z-score method [30] and the PyOD package [31], both renowned tools in the field of anomaly detection. After a meticulous examination, the outliers are identified and excluded according to the three following distinct aspects:

1) **Transition temperature**: The average transition temperature of remaining compounds ranges from 0.0005 to 250 K. Using the abovementioned techniques, 10 superconducting materials with average transition temperatures outside the 0.01-136 K range are identified as outliers and removed. Figure 2 illustrates the $T_c$ distribution of the few superconducting material families.

2) **Number of elements**: Figure 3 shows the number of compounds according to the number of constituent elements. A subset of compounds with one, eight, and ten elements are identified as outliers and subsequently removed, resulting in the elimination of 81 superconducting compounds.

3) **The summation of subscripts**: Implementing the abovementioned techniques reveals that six compounds exhibited subscript summations exceeding 100 that are subsequently removed as outliers.

Our meticulous data-cleaning procedures yield a refined dataset, called DataG dataset, containing 13,022 compounds.

## 3. Computational Methods

### 3.1. Machine learning Algorithm

In this study, we use the CatBoost algorithm as a machine learning ensemble technique based on Gradient Boosted Decision Trees (GBDT) proposed by Yandex Company. GBDT is an efficient tool for solving regression and classification issues in big data sets. CatBoost is a Decision Tree based algorithm and open-source implementation for supervised machine learning that involves two innovations: Ordered Target Statistics and Ordered Boosting. Researchers have successfully employed CatBoost for machine learning investigations incorporating Big Data since its launch in late 2018. Numerous applications have been reported for CatBoost in various fields, including astronomy, finance, medicine, biology, electrical utilities fraud, meteorology, psychology, traffic engineering, cyber-security, biochemistry, and marketing [32]. However, the application of CatBoost has not yet been reported for predicting superconducting transition temperatures. This study uses the algorithm to find if it can efficiently identify relationships and patterns between features and $T_c$. We show that through the creation of atomic features for superconducting material, CatBoost algorithm provides a model with very good accuracy.

### 3.2. Generating the Feature Space

After preprocessing the data set, we must extract atomic features in a "data representation" procedure. There are two main approaches for representing compounds: The first is based on chemical formulas, and the second on crystal structure [23]. The atomic features are generated for superconducting materials using the first approach for our purposes.

In fact, machine learning algorithms recognize a compound by its characteristics, *i.e.* the identifier and characteristic of a compound are the features that are consider for the compound. This process is called data representation. Figure 4 shows how to calculate atomic features.

We design and develop the Python language package called Jabir to generate 322 atomic features for each compound of all types, including superconducting materials. The package calculates eight statistical relationships (e.g., variance and mean) for each physical feature (e.g., magnetic moment) based on the three components of Element, Subscript, and Fraction. Figure 5 depicts the workflow of the feature-generating by Jabir.

For illustration, consider the compound $Mg_{0.9}Fe_{0.1}B_2$ composed of three elements. The subscripts are 0.9, 0.1, and 2, while the fractions of the elements in the compound, obtained from equation 1, are 0.3, 0.033, and 0.666, respectively.

$$Fraction = \frac{Subscript}{\sum Subscript} \qquad (1)$$

As mentioned, the atomic features are generated based on the three components of Element, Subscript, and Fraction. The fraction-based atomic features are multiplied by the fraction of the element in the compound.

Similarly, the subscript-based atomic features are multiplied by the subscript of that element in the compound. However, the atomic features based on element (Element-based) are based solely on elemental values; in other words, the elemental value is multiplied by one, ignoring the related fraction or subscript. It should also be noted that the Jabir package solely calculates the element-based atomic features for the four Ionic Radius, Vander Waals Radius, Period Number and Group Number features because the two subscript-based and fraction-based ones are meaningless for these features. Table 1 briefly explains the process used for calculating the mean thermal conductivity of the $Mg_{0.9}Fe_{0.1}B_2$ compound, for instance. To learn more about Jabir's features, see the Supplementary Information; we have explained briefly all the 30 most significant features depicted in Figure 7.

At first glance, it seems fraction-based and subscript-based are the same thing, and we should choose one. However, the fraction of some elements in some compounds consisting of the same elements can be equal while they have different subscripts. The subscript-based must also be considered in atomic features space to account for this difference. For instance, the compounds $Y_1Fe_2Si_2$ and $Y_2Fe_3Si_5$ have identical fractions of "Y" (namely, 0.2), while it is rational to think that this same element has different effects in these two compounds. Clearly, among the three types of atomic features, only the subscript-based one accounts for this difference, demonstrating the reason why the subscript-based feature must be used.

### 3.3. Feature Selection

Feature selection methods are utilized to determine the best feature subset. Some advantages of feature selection include reduced overfitting, improved accuracy, reduced training time, simplified model design, faster convergence, enhanced generalization, and improved robustness to noise [33-36]. Feature selection methods help pick out the subset of attributes most relevant to the $T_c$ of superconducting materials. Generally, for a feature space of N features, there are $2^N$ subsets of features. For example, the current feature space with 322 features leads us to select a subset among $2^{322} \approx 8.54 \times 10^{96}$ feature subsets. Due to the enormous number of subsets, we need sophisticated methods to overcome computational costs. There are generally four general methods for selecting a subset of features: filter, wrapper, embedded, and hybrid [33, 35, 37]. In this study, the various feature selection techniques are tested carefully and evaluated for their efficiency and effectiveness against such evaluation criteria as the coefficient of determination ($R^2$), Eq. 2, and root mean square error (RMSE), Eq. 3 [23]. Finally, we developed a novel and innovative hybrid method. This method has been published in the form of a Python package called Soraya.

$$R^2 = 1 - \frac{\sum_{i=1}^{n}(y_i - \hat{y}_i)^2}{\sum_{i=1}^{n}(y_i - \bar{y})^2} \quad (2)$$

$$RMSE = \sqrt{\frac{1}{n}\sum_{i=1}^{n}(y_i - \hat{y}_i)^2} \quad (3)$$

Using the proposed feature selection technique, 30 of the most important features generated by the Jabir package for DataG are selected. The results of these comparisons are reported in Table 2.

The proposed innovative hybrid method for selecting the best subset of feature space calculates the two-by-two correlation of all 322 features in its first step. In the second step, all the features with absolute Pearson correlation criterion values greater than 0.80 are grouped into distinct clusters. The Pearson correlation criterion is a parametric statistical method that allows to determine the existence or absence of a linear relationship between two quantitative variables [38]. This step categorizes 304 features into 62 clusters and retains 18 features with correlations less than 0.80 for the following steps. (Each cluster contains a varying number of features. For example, one cluster may consist of only two correlated features, while another may comprise five. Therefore, the rationale behind Soraya's decision to group 304 features into 62 clusters is based on the characteristics of the features presented in the dataset.). This step aims to retain the most important features from each cluster and eliminate the remainder as redundant. Features with a correlation greater than 0.8 mean that they have very strong correlation [39]. The objective of feature selection is to identify a subset of the original features from a provided dataset by eliminating irrelevant and redundant features [40]. Furthermore, as the feature space dimensions decrease, the learning model's accuracy will increase [41]. We should keep only one of those features which they have very strong correlation and remove the others, because they are redundant features that do not provide any new information [42].

In the third step, the learning process of the model is performed independently for each cluster, with the most significant feature in each cluster being selected and the others eliminated. As a result of grouping the features into 62 distinct clusters, 62 features remain in this step. Once the duplicate features have been deleted, 55 features remain.

In the fourth step, 18 features are added to the 55 features obtained above. Subsequently, the SHAP (SHapley Additive exPlanations) method, which is based on the game theory for explaining the output of machine learning models [43, 44], is employed to sort the 73 features according to the significance level. In this step, the SHAP method acts as a filter method.

In the fifth step, the 5 most significant features, as identified and sorted by the SHAP method in the previous step, are initially selected. Using the forward selection (wrapper method), the remaining 68 (*i.e.*, 18+55−5) features are added one by one to the 5 features, until 30 features are selected from among the most significant ones. (The 30 most significant features give the highest accuracy for the model; The Soraya package is designed in such a way that it shows the amount of accuracy with the addition of each feature.) The steps outlined above are depicted in Figure 6.

The DataG dataset, which contains 13022 superconducting materials, comprises 83 elements from the periodic table. As a result, 83 columns are created in which the fractions of the constituent elements for each compound are recorded. This process is illustrated in Table 3. Finally, we add this feature vector to the previously selected features to make a final feature space with 113 (83+30) dimensions.

## 4. Results

### 4.1. Identifying Key Features

During the feature selection process in section 3.3, we employed an innovative hybrid technique called the "Soraya package" to pick 30 of the most significant features. Subsequently, in Figure 7, we sorted these selected features using the capability of the CatBoost algorithm. In the Supplementary Information, we have explained those features depicted in Figure 7.

Across various studies [4, 38, 45], including the current study, researchers have discovered that the thermal conductivity stands out as the most important feature among different features in determining the $T_c$ of superconducting materials. Theoretically, the thermal conductivity of superconductors provides significant clues about the nature of their charge carriers, phonons, and the scattering processes occurring between them [46]. Thermal conductivity refers to the ability of a material to conduct heat. The significance of thermal conductivity is directly connected to the concentration of particles capable of transferring heat [45, 47]. The concentration of the superconducting particles ($n_s$) is related to a characteristic length describing the superconducting state, namely, the London penetration depth ($\lambda$), $\lambda^2 = \frac{m}{q^2 n_s \mu_0}$, where m, q, $n_s$ are mass, charge and concentration of superconducting particles respectively and $\mu_0$ is magnetic constant [38, 48]. The transition temperature of a superconductor is associated with both the London penetration depth and the coherence length. In other words, the formation and destruction of the superconducting state is related to the London penetration depth and the coherence length [38, 45, 48]. On the other hand, the results of this study and other studies [4, 38, 45] show that the superconducting transition temperature has a strong correlation with the thermal conductivity; Among the 322 features, the range of thermal conductivity has the strong correlation (0.68) with the $T_c$; see Fig. 7. Then, it could be concluded that the results of this research are consistent with the results of theoretical works.

### 4.2. Predicting the superconducting materials' $T_c$ values

For the DataG dataset, which contained 13022 superconducting materials, 90% of the dataset is allocated to the training dataset and 10% to the test one (*i.e.*, 1303 compounds). Using the model created during training, the CatBoost algorithm predicted a $T_c$ value for every 1303 compounds in the test dataset (Figure 8). The $R^2$ and RMSE evaluation criteria are 0.952 and 6.45, respectively, superior to those reported in the literature. Table 4 compares the values for the $R^2$, RMSE, and MAE evaluation criteria obtained in the present study and those reported elsewhere. The model proposed here yields $R^2$, RMSE, and MAE values superior to those previously reported.

Furthermore, the procedure employed for DataG (namely, creating new features, selecting features, tuning hyperparameters, etc.) is also applied to DataS, DataK, and DataH datasets, which led to improved evaluation criteria (Table 4).

The model thus developed is subsequently used to predict the $T_c$ values for $SmFeAsO_{0.8}F_{0.2}$, $SmFeAsO_{0.7}F_{0.3}$ and three new Iron-based superconducting materials not included in the data set and for which no $T_c$ values had yet been reported. As shown in Table 5, the $T_c$ value of the main compound increases as the Fluorine element increases to an optimal doping content. This increase in $T_c$ aligns with the experimental results. Table 5 also indicates that we can play around with elements (e.g., substitution, changing contribution) and find trends for increasing $T_c$ for a specific compound, which can help material scientists to design high-temperature superconductors. It should be emphasized that none of the compounds mentioned in Table 5 are included in the SuperCon dataset.

Moreover, the model was used to predict the $T_c$ values for a few superconducting compounds not included in the DataG but for which $T_c$ values had been previously reported in the literature. The results are provided in Table 6 for comparison. Clearly, a great agreement can be observed between the two $T_c$ values obtained by experiments and by the machine learning method used in this study.

## 5. Conclusion

In the realm of materials science, artificial intelligence stands as a powerful tool for predicting material properties. In this study, the CatBoost algorithm was employed to predict the $T_c$ values of superconducting materials, marking a novel approach. For this purpose, data pre-processing of the SuperCon dataset was accomplished as a significant step in data science to develop a new dataset called DataG containing 13022 superconducting compounds. Also, a new Jabir package capable of generating 322 atomic descriptors was designed and developed. Comparisons revealed the superiority of the atomic features generated by Jabir over those generated by such previous ones as the Magpie package. Furthermore, an innovative hybrid technique was developed as the feature selection method (Soraya package). In order to design and develop Jabir and Soraya packages, we applied novel ideas and innovative approaches, such as: i) using new and diverse physical atomic features in the Jabir package and considering three different states (Elemental, Subscript, Fraction) in order to calculate the atomic features of each compound and ii) using an innovative hybrid technique in Soraya package, removing features that are highly correlated with each other (removing redundant features) and using SHAP's technique to select the most important features and finally using the forward method to adding the most important features. The contributions of the study led to optimized evaluation values ($R^2$, RMSE, MAE) of DataH, DataS, and DataK datasets without the need for any data pre-processing. The present study's results indicate that the procedure of selecting the most important descriptors significantly impacts predicting superconducting materials' $T_c$ values. Finally, the development of a novel web application was a pioneering contribution to the field for predicting and determining the $T_c$ of superconducting materials.

**Data availability**

The dataset (DataG), which is prepared after various steps of data pre-processing on the SuperCon dataset, is available at the following address.

https://github.com/Gashmard/DataG_13022_superconducting_materials

**Code availability**

The developed packages (Jabir and Soraya) and the web application are accessible at the following URLs. Web application: https://supercon-tc.iut.ac.ir/

jabir package: https://pypi.org/project/jabir/

soraya package: https://pypi.org/project/soraya/

jabir package on Github: https://github.com/Gashmard/jabir

Soraya package on Github: https://github.com/Gashmard/Soraya

**FIGURES**

Figure 1: Four paradigms of materials science from the beginning to the current.

Figure 2: Distribution of transition temperature of superconducting materials.

Figure 3: The number of superconducting compounds based on the number of their constituent elements.

Figure 4: Atomic feature generation workflow for a compound.

Figure 5: Jabir package generates 322 atomic features for each compound, including eight statistical relations for 12 physical features based on the three components of Elements, Subscript, and Fraction. Also, for the four Ionic Radius, Vander Waals Radius, Period Number and Group Number features, only the Element component is calculated because the two others, subscript-based and fraction-based ones, are meaningless for these features. Therefore, 320 features have been generated, and finally, 2 more features are added to them: the number of constituent elements of each compound and the sum of their subscripts. In short, there are 322 atomic features for each compound ($12 \times 3 \times 8 + 4 \times 1 \times 8 + 2 = 322$).

Figure 6: Workflow related to the innovative hybrid method (Soraya Package). Through a 5-step process, this technique can efficiently identify the most significant features in the dataset while eliminating any redundant features.

Figure 7: Using Soraya package, 30 of the most significant features were selected from among 322 features, and subsequently, the CatBoost algorithm was employed to sort these selected 30 features.

Figure 8: Comparison of predicted $T_c$ of 1303 superconducting compounds by machine learning model according to experimental $T_c$.

Table 1: Calculation of the mean thermal conductivity for $Mg_{0.9}Fe_{0.1}B_2$ compound.

| Atom | Subscript | Fraction | Element | Thermal conductivity (W.K$^{-1}$.m$^{-1}$) | Subscript Thermal conductivity | Fraction Thermal conductivity | Element Thermal conductivity |
|---|---|---|---|---|---|---|---|
| $Mg_{0.9}$ | 0.9 | 0.3 | 1 | 160 | 160 × 0.9=144 | 0.3 × 160=48 | 160 |
| $Fe_{0.1}$ | 0.1 | 0.033 | 1 | 80 | 80 × 0.1=8 | 0.033 × 80=2.64 | 80 |
| $B_2$ | 2 | 0.66 | 1 | 27 | 27 × 2=54 | 0.66 × 27=17.82 | 27 |
| mean | - | - | - | - | 68.66 | 22.82 | 89 |

Table 2: Evaluation and comparison of various feature selection techniques. There are four general methods for selecting the most important features, each encompassing multiple techniques. Here, each technique selects 30 of the most significant features and subsequently, their performance is evaluated.

| Method | Technique | N_Features | $R^2$_test | RMSE_test |
|---|---|---|---|---|
| Embedded | Lasso Regression | 30 | 0.937 | 7.31 |
| Embedded | Ridge Regression | 30 | 0.929 | 7.78 |
| Embedded | Elastic Net | 30 | 0.937 | 7.31 |
| Wrapper | Forward | 30 | 0.931 | 7.68 |
| Wrapper | Backward | 30 | 0.929 | 7.77 |
| Filter | shuffling | 30 | 0.933 | 7.55 |
| Filter | Pearson correlation | 30 | 0.896 | 9.42 |
| Filter | Permutation | 30 | 0.940 | 7.14 |
| Filter | Decision Tree | 30 | 0.935 | 7.46 |
| Filter | Random Forest | 30 | 0.937 | 7.34 |
| Filter | XGBoost | 30 | 0.916 | 8.46 |
| Filter | CatBoost | 30 | 0.941 | 7.09 |
| Filter | SHAP Value | 30 | 0.939 | 7.21 |
| Filter | Featurewiz Package | 30 | 0.937 | 7.35 |
| Filter | Mutual Information | 30 | 0.926 | 7.96 |
| Filter | Jostar Package | 30 | 0.935 | 7.48 |
| Filter | Genetic Algorithm | 30 | 0.934 | 7.49 |
| Hybrid | **Soraya** | 30 | **0.947** | **6.8** |

Table 3: Ba$_{0.8}$Fe$_2$Se$_2$ is represented by an array of 83 columns, with each column containing the fraction of constituent elements present in Ba$_{0.8}$Fe$_2$Se$_2$.

| Compound | Ba$_{0.8}$Fe$_2$Se$_2$ |
|---|---|
| H | 0 |
| He | 0 |
| Li | 0 |
| … | 0 |
| Fe | 0.417 |
| … | 0 |
| Se | 0.417 |
| … | 0 |
| Ba | 0.167 |

Table 4: Evaluation values for predicting the $T_c$ of superconducting materials related to the current research and other works. (GBDT: Gradient Boosting Decision tree), (CNN: Convolutional Neural Network), (ConvGBDT: Convolutional Gradient Boosting Decision tree), (HNN: Hybrid Neural Network), (CGCNN: Crystal Graph Convolutional Neural Network).

| Authors | Algorithm | Features | Dataset | MAE (K) | RMSE (K) | $R^2$ |
|---|---|---|---|---|---|---|
| K.Hamidieh [4] | XGBoost | 80 atomic features | DataH(21263) | - | 9.5 | 0.92 |
| V.Stanev [16] | Random Forest | Magpie(145 features) | 16415 | - | - | 0.885 |
| Y.Dan [23] | GBDT | Magpie(132 features) | DataS(6200) | 6.92 | 10.92 | 0.873 |
| Y.Dan [23] | CNN | Magpie(132 features) | DataS | 7.88 | 12.02 | 0.831 |
| Y.Dan [23] | ConvGBDT | Magpie(132 features) | DataS | 5.51 | 8.93 | 0.907 |
| Y.Dan [23] | ConvGBDT | Magpie(132 features) | DataH | 4.65 | 8.69 | 0.937 |
| Y.Dan [23] | ConvGBDT | Magpie(132 features) | DataK(12700) | 4.74 | 8.83 | 0.931 |
| S.Li [11] | HNN | Magpie(132 features) | 12413 | 5.023 | - | 0.899 |
| S.Li [11] | Random Forest | Magpie(132 features) | 12413 | 5.096 | - | 0.88 |
| B.Roter [24] | Bagged Tree Al | chemical composition matrix | 30000 | - | 8.91 | 0.93 |
| T.Konno [3] | Deep Learning | 7×32×4 Reading periodic table | DataK | - | - | 0.92 |
| M.Quinn [25] | CGCNN | crystal-graph representation | ≈ 15000 | 5.6 | - | 0.92 |
| **Present work** | **CatBoost** | **Jabir** | **DataG(13022)** | **3.54** | **6.45** | **0.952** |
| Present work | CatBoost | **Jabir** | DataH | 4.62 | 8.18 | 0.942 |
| Present work | CatBoost | **Jabir** | DataK | 4.93 | 8.83 | 0.937 |
| Present work | CatBoost | **Jabir** | DataS | 5.37 | 8.69 | 0.914 |

Table 5: Prediction of $T_c$ of three new Iron-based superconducting compounds using machine learning model. By substituting an element instead of Samarium, the value of $T_c$ changes.

| Compound | $T_c$ ML (K) | $T_c$ Exp (K) |
|---|---|---|
| $SmFeAsO_{0.8}F_{0.2}$ | 46.42 | $\approx 50$ [49] |
| $SmFeAsO_{0.7}F_{0.3}$ | 49.46 | $\approx 53$ [50] |
| $Sm_{0.8}Rh_{0.2}FeAsO_{0.8}F_{0.2}$ | 27.28 | - |
| $Sm_{0.8}Ir_{0.2}FeAsO_{0.8}F_{0.2}$ | 29.27 | - |
| $Sm_{0.8}In_{0.2}FeAsO_{0.8}F_{0.2}$ | 39.84 | - |

Table 6: The machine learning model developed in this study accurately predicts the $T_c$ of several superconducting compounds, none of which are present in the original SuperCon dataset.

| Compound | $T_c$ ML (K) (**DataG**) | $T_c$ Exp (K) |
|---|---|---|
| $SrAl_2Si_2$ [51] | 3.37 | 4.6 |
| $FeTa_{0.02}Se$ [52] | 10.29 | 8 |
| $FeNi_{0.02}Se$ [52] | 7.81 | 5 |
| $LaFeO_{0.9}F_{0.1}$ [53] | 18.52 | 26 |
| $Ba_{0.6}K_{0.3}Fe_2As_2$ [52] | 22.82 | 38.6 |
| $Ba_{0.29}K_{0.71}Fe_2As_2$ [52] | 18.52 | 14.6 |
| $NdFeAsO_{0.89}F_{0.11}$ [51] | 48.06 | 48 |
| $NdFeAsO_{0.82}F_{0.18}$ [53] | 48.28 | 51 |
| $SmFeAsO_{0.8}H_{0.2}$ [51] | 53.37 | 55 |
| $Ca_{0.83}La_{0.17}Fe_2As_{1.88}P_{0.12}$ [54] | 38.01 | 45 |

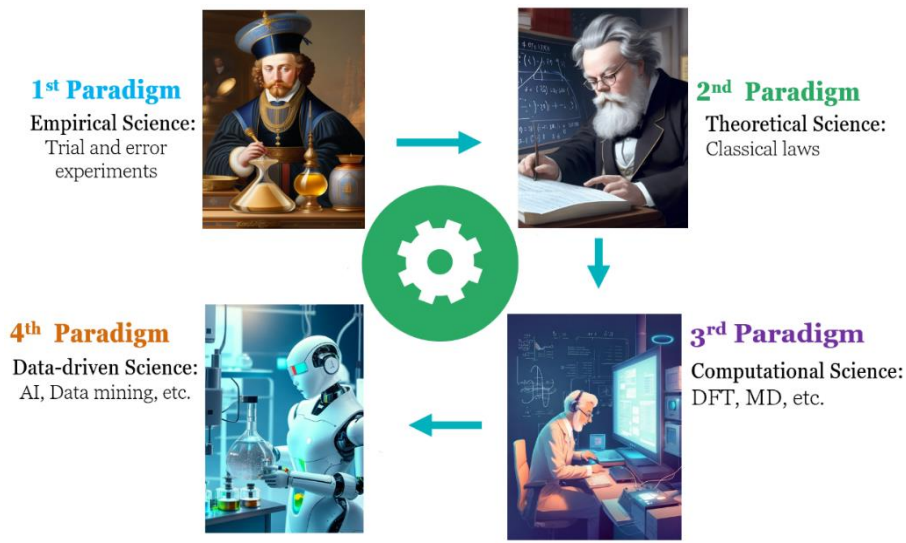

Figure 1: Four paradigms of materials science from the beginning to the current.

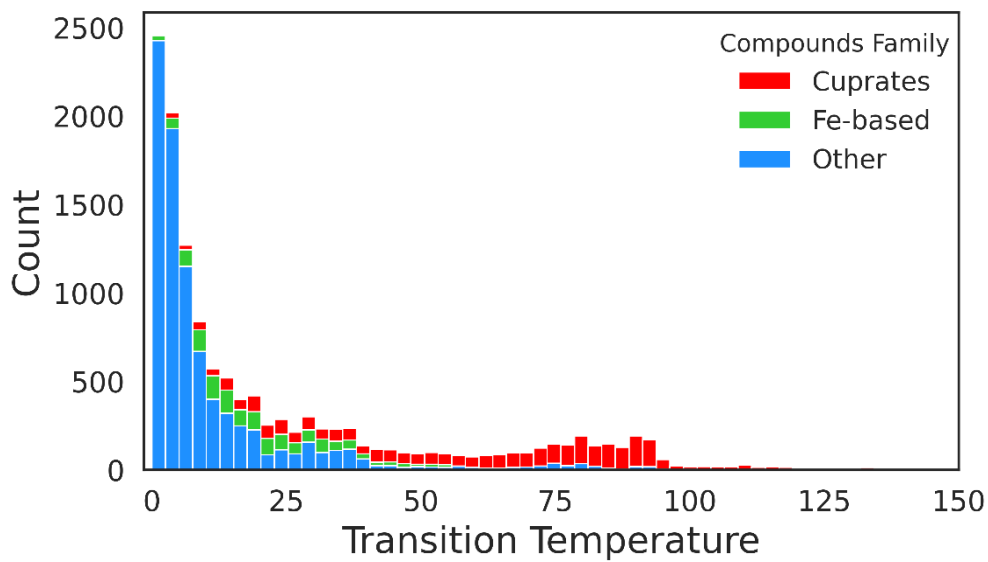

Figure 2: Distribution of transition temperature of superconducting materials.

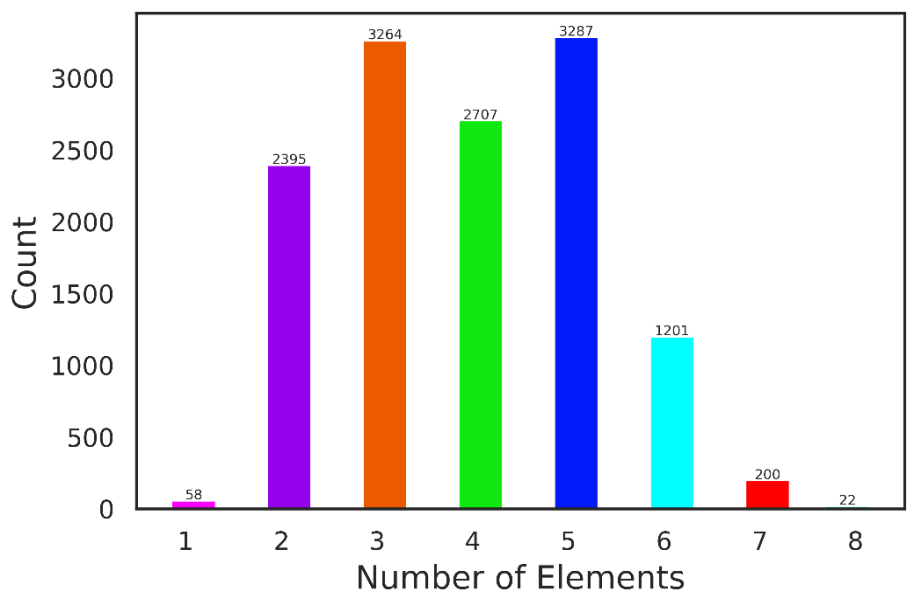

Figure 3: The number of superconducting compounds based on the number of their constituent elements.

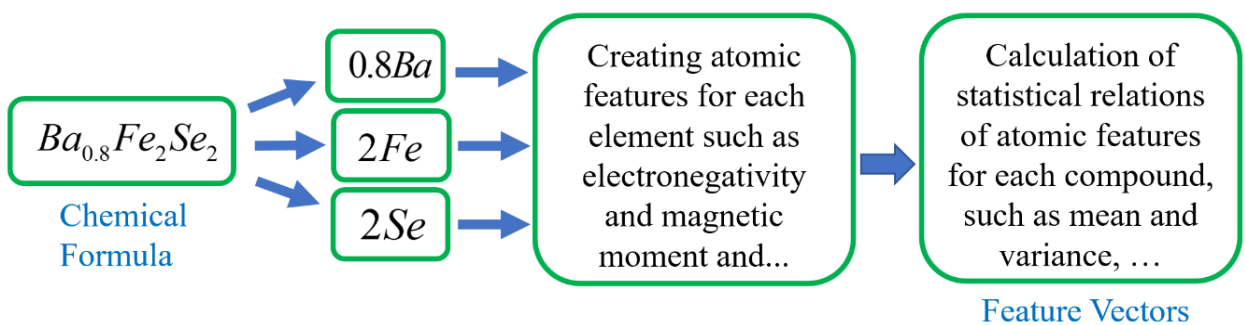

Figure 4: Atomic feature generation workflow for a compound.

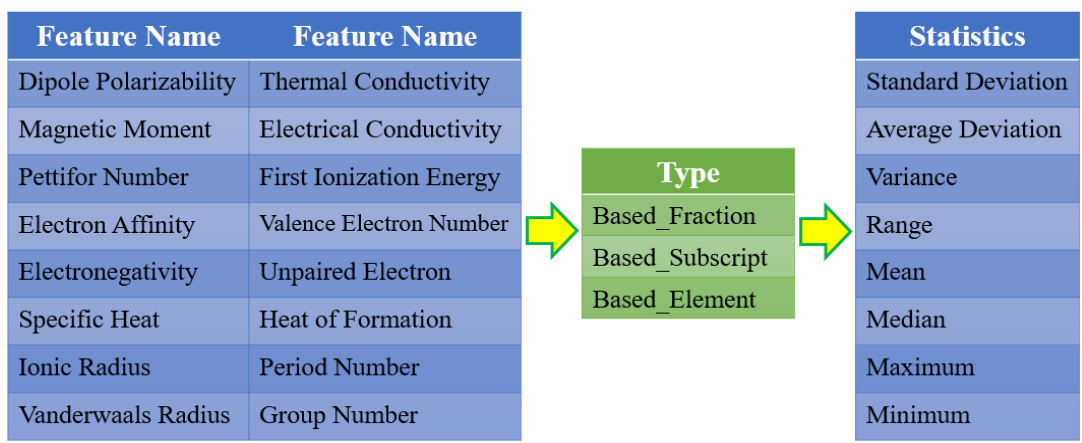

Figure 5: Jabir package generates 322 atomic features for each compound, including eight statistical relations for 12 physical features based on the three components of Elements, Subscript, and Fraction. Also, for the four Ionic Radius, Vander Waals Radius, Period Number and Group Number features, only the Element component is calculated because the two others, subscript-based and fraction-based ones, are meaningless for these features. Therefore, 320 features have been generated, and finally, 2 more features are added to them: the number of constituent elements of each compound and the sum of their subscripts. In short, there are 322 atomic features for each compound ($12 \times 3 \times 8 + 4 \times 1 \times 8 + 2 = 322$).

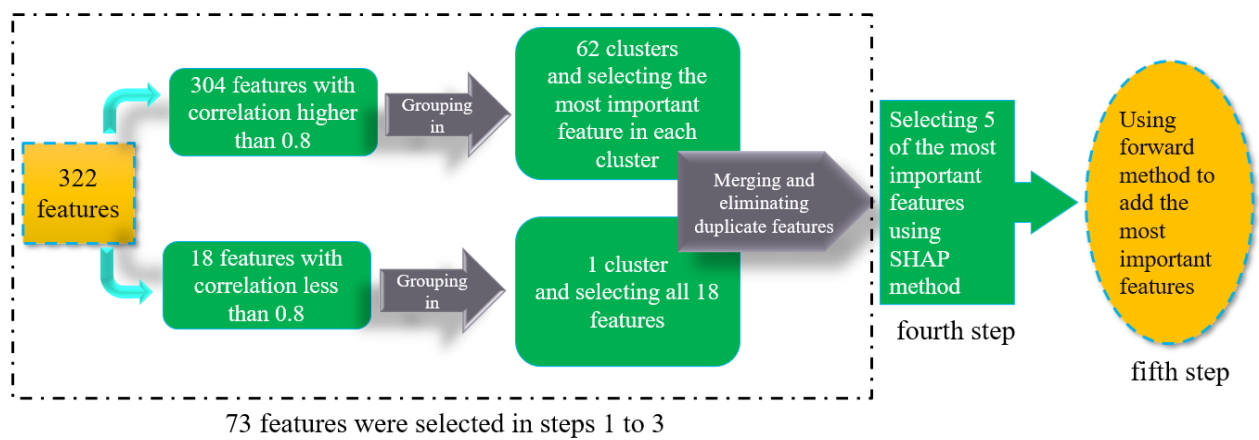

Figure 6: Workflow related to the innovative hybrid method (Soraya Package). Through a 5-step process, this technique can efficiently identify the most significant features in the dataset while eliminating any redundant features.

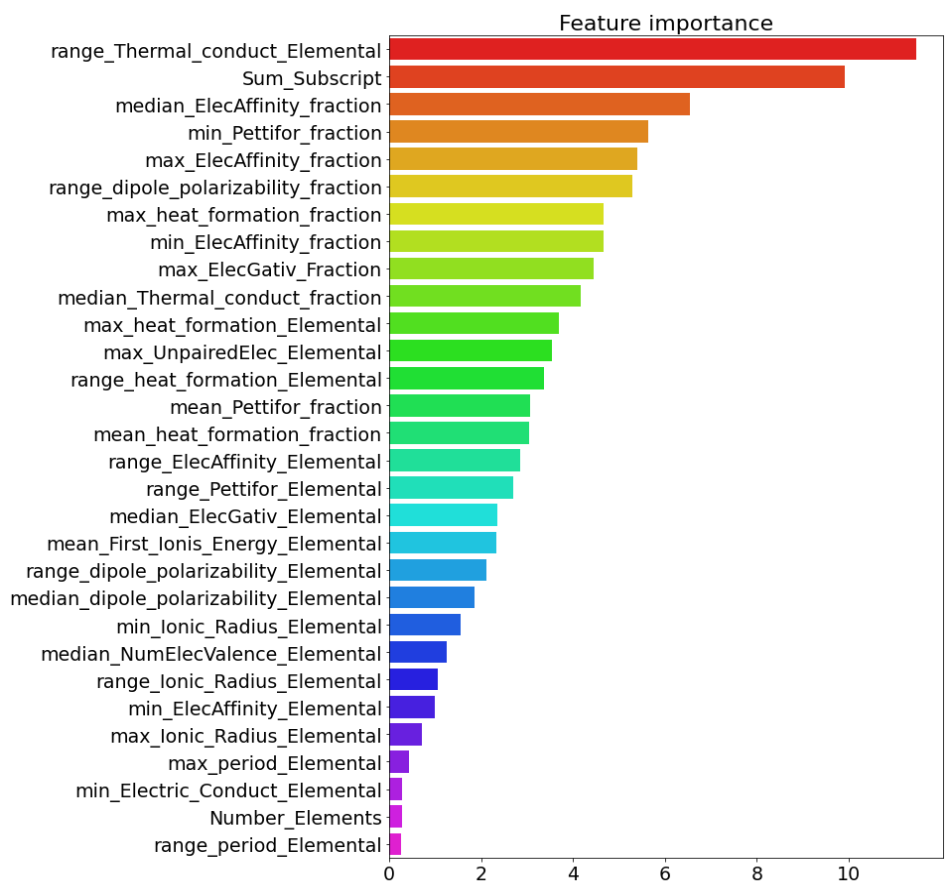

Figure 7: Using Soraya package, 30 of the most significant features were selected from among 322 features, and subsequently, the CatBoost algorithm was employed to sort these selected 30 features.

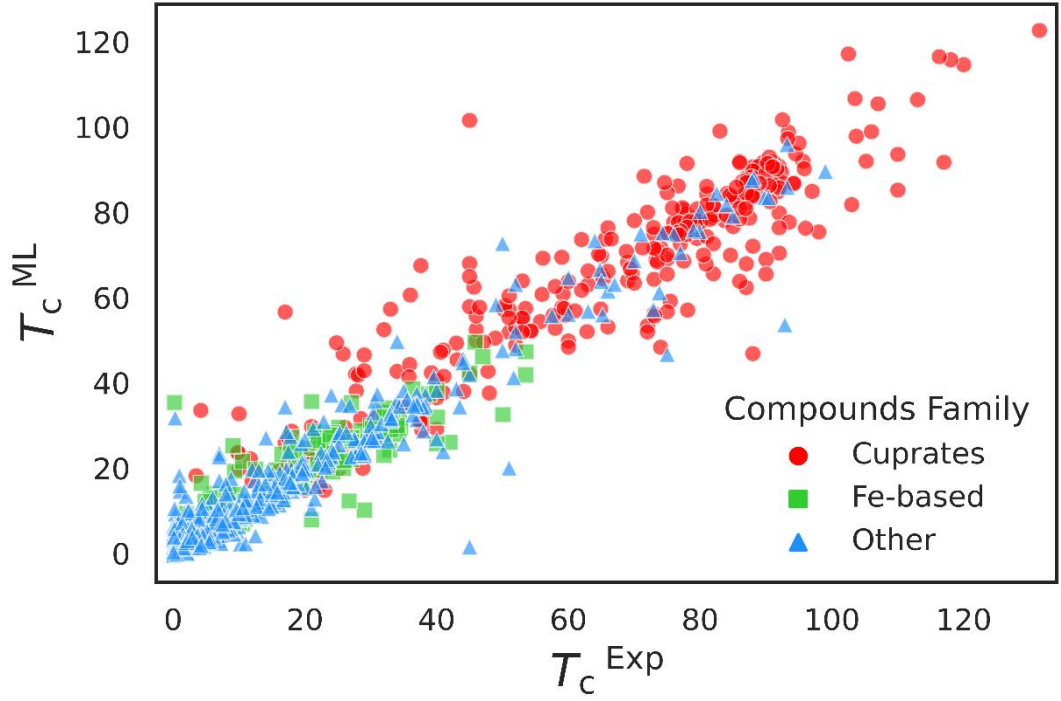

Figure 8: Comparison of predicted $T_c$ of 1303 superconducting compounds by machine learning model according to experimental $T_c$.